\documentclass[12pt]{article}

\usepackage{graphics,psfrag}
\usepackage{amsthm,amssymb,epsfig,amsmath,euscript,array,cite}

\newcommand{\be}{\begin{equation}}
\newcommand{\ee}{\end{equation}}
\def\bea{\begin{eqnarray}}
\def\eea{\end{eqnarray}}
\newcommand{\eq}[1]{(\ref{#1})}
\def\nn{\nonumber}

\newcommand{\beq}{\begin{equation}}
\newcommand{\eeq}{\end{equation}}
\newcommand{\ben}{\begin{eqnarray}}
\newcommand{\een}{\end{eqnarray}}
\newcommand{\bes}{\begin{subequations}}
\newcommand{\ees}{\end{subequations}}
\newcommand{\blg}{\begin{align}}
\newcommand{\elg}{\end{align}}

\newcommand{\cN}{{\cal N}}

\newcommand{\startappendix}{
\setcounter{section}{0}
\renewcommand{\thesection}{\Alph{section}}}

\newcommand{\Appendix}[1]{
\refstepcounter{section}
\begin{flushleft}
{\large\bf Appendix \thesection: #1}
\end{flushleft}}

\def\one{\mbox{1 \kern-.59em {\rm l}}}

\def\a{\alpha}      
\def\b{\beta}       
    
\def\d{\delta}  \def\D{\Delta}

\def\k{\kappa}
\def\l{\lambda} 
 
\def\o{\omega}
 
\def\r{\rho}
\def\s{\sigma}  \def\S{\Sigma}
\def\t{\tau}
\def\th{\theta}

 \def\cE{{\cal E}} 
  
\def\cJ{{\cal J}}  
 \def\cN{{\cal N}}

 \def\cZ{{\cal Z}}

\begin{document}

\hfill{DCPT-09/03}

\vspace{20pt}

\begin{center}

{\Large \bf
Semi-classical Strings in Sasaki-Einstein Manifolds
}
\vspace{20pt}

{\bf
 Dimitrios Giataganas
}

{\em
Centre for Particle Theory and Department of Mathematics,\\
Durham University, South Road,
Durham, DH1 3LE, UK.
}

{\small \sffamily 
dimitrios.giataganas@durham.ac.uk
}

\vspace{30pt}
{\bf Abstract}
\end{center}
We find point-like and classical string solutions on the $AdS_5\times X^5$, where $X^5$ are the 5-dimensional Sasaki-Einstein manifolds $Y^{p,q}$ and $L^{p,q,r}$. The number of acceptable solutions is limited drastically in order to satisfy the constraints on the parameters and coordinates of the manifolds. The energy of the solutions depends on the parameters of the Sasaki-Einstein manifolds and on the conserved momenta transcendentally. A discussion on BPS solutions is presented as well.

\setcounter{page}0
\newpage

\section{Introduction}

The idea that large $N$ gauge theories may have a string theory description was proposed a long time ago \cite{tHooft1}, but the first concrete proposal was given by Maldacena with the conjectured AdS/CFT correspondence \cite{maldacena1,witten1,gubser1}. The first and most studied model was the duality between the large $N$ limit (where $\l=g_{YM}^2 N$ is kept fixed), maximally supersymmetric, $\cN=4$ Yang Mills theory in four dimensions with gauge group $SU(N)$, and type IIB supergravity on $AdS_5\times S^5$ with N units of $RR$ 5-form flux through $S^5$. The duality is so useful as the radius of curvature $R$ of the $AdS_5$ space, and the 't Hooft coupling $\l$, are related through $\l=R^4/l_{str}^4$, where $l_{str}$ is the string length. When the curvature radius $R$ is much  bigger than the length of the string, the worldsheet theory becomes classical and can be studied. In this regime, $\l$ approaches infinity and  the dual gauge theory is strongly coupled and not under best control, which makes the calculations on the gravity side valuable and the correspondence useful. On the other hand, this fact makes it difficult to realize the concrete connection between the string theory and the gauge theory beyond the supergravity approximation.

Work in this direction was made in \cite{nastase1}, where certain gauge theory operators with large R-symmetry charge was proposed to be dual to all type IIB string states in a RR-charged pp-wave background \cite{metsaev1}, which is a Penrose limit of $AdS_5\times S^5$ \cite{blau1}. A generalization came in \cite{gubser2}, where with the use of classical solitons and an appropriate quantization of the string theory, it has been shown that there are some semiclassical limits where the the string/gauge duality can be reliable. For short strings, where one can approximate $AdS_5$ by a flat metric near the center, the leading closed string trajectory is reproduced. On the other hand for long strings, the strings feel the metric near the boundary of $AdS$, and reproduce the logarithmic behavior of the scaling of the wavefunction renormalization and hence the anomalous dimension for operators with ''dimension minus spin''  equal to two. Based on this work, a further generalization was made by considering multi-spin string states \cite{frolov1,frolov2}. At this time, many papers were published  with the aim of finding new close string solutions, for example see \cite{tseytlin1} and references inside.

Another significant step was the identification of the one loop scalar dilatation operator with the Hamiltonian of integrable $SO(6)$ spin chains \cite{minahan1}. This has brought impressive quantitative agreements between the energy of certain string solutions and the anomalous dimensions for very long operators \cite{beisert1}. Almost one year later, it was shown \cite{kruczenski1} that the spin chain in a certain subsector, in the limit of a large number of sites, can be described by a sigma model which agrees with the sigma model obtained from the rotating string in the appropriate limit.

All these results lead to further study with the aim to extend the above analysis to non-conformal theories or conformal theories with less supersymmetries. Here we concentrate on the latter and mention briefly that a famous effort in this direction was made possible with the use of the Leigh-Strassler \cite{leigh1} marginal deformation. Lunin and Maldacena, using a special case of this deformation on the original gauge/gravity duality, and taking advantage of the global $U(1)\times U(1)$ symmetry that the deformed theory and background should respect, found a new duality between $\cN=1$ super Yang-Mills theory and a deformed background \cite{lunin1}. This development gave new grounds for further testing and analyzing the $AdS/CFT$ duality. Some characteristic examples including instantons, mesons, gravitons, Wilson loop calculations are in \cite{frolov3}. Equally important was the extension of the duality by Frolov \cite{frolov4} to non-supersymmetric theories. In some cases this made it possible to examine whether or not certain properties of field theories are related  to supersymmetry \cite{giataganas2}.

Another famous gauge/gravity correspondence for theories with less supersymmetries, is with the use  of a class of backgrounds with at least $\cN=1$ supersymmetry, which are type IIB and with the form $AdS_5\times X^5$, where $X^5$ is a Sasaki-Einstein manifold and are dual to superconformal gauge theories called quivers.

A Sasakian manifold is a Riemannian manifold whose metric cone is K\"{a}hler. A K\"{a}hler manifold is a Hermitean manifold $(M,g)$, say of dimension $m$, whose K\"{a}hler form $\Omega$, defined by
\be
\Omega_p(X,Y)=g_p(J_pX,Y)\qquad X,Y \in T_p M,
\ee
where $J_p$ is the almost complex structure, it is closed. This form can be used to prove that a complex manifold is orientable, since the real $2m$ form $\Omega\wedge...\wedge\Omega$ vanishes nowhere and it serves as a volume element. Additionally, if the manifold $M$ is compact and admits a Ricci flat metric, then its first Chern class must vanish and the manifold is called Calabi-Yau.  The Sasaki-Einstein manifolds, are manifolds whose metric cones are Ricci flat and K\"{a}hler.

The first non-trivial example in the AdS/CFT correspondence with the use of these manifolds was made in the case of the manifold $T^{1,1}$ \cite{klebanov1}. Significant progress has been made in Sasaki-Einstein backgrounds and their dual field theories, almost a year before the Lunin-Maldacena correspondence was formulated,  when it was  found that for five-dimensional Sasaki-Einstein manifolds $Y$,  there is an infinite family of inhomogeneous metrics on $Y^{p,q}\cong S^2\times S^3$, which is characterised by relatively prime positive integers $p,\,q$ with $0<q<p$ \cite{gauntlett1,gauntlett2,gauntlett3}. In this case, there is an effective action of a torus $T^3\cong U(1)^3$ on the $C(Y^{p,q})$ which preserves the symplectic form on it and the metric, since it is an isometry. In \cite{martelli1} there is an extensive discussion on the geometric features of these manifolds and in a paper \cite{benvenuti1} that followed some days after, the superconformal quiver gauge theories dual to type IIB string theory on $AdS_5\times Y^{p,q}$ was proposed.

The  spaces $Y^{p,q}$ are of cohomogeneity one, but the correspondence can be generalized to spaces with cohomogeneity two, called $L^{p,q,r}$ spaces \cite{cvetic1}. These are characterized by the relative positive coprime integers $p,\, q$ and $r$ with $0<p\leq q,\,0 < r < p+q$ and with $p,\,q$ to be coprime to $s=p+q-r$ and have isometry $U(1)\times U(1)\times U(1)$. The metrics $Y^{p,q}$ are a special case of $L^{p,q,r}$ where $p+q=2r$.

Moreover, like all theories with at least a $U(1)\times U(1)$ global symmetry, the toric quiver gauge theories and their gravity dual theories admit $\beta$ deformations \cite{benvenuti2}, which also give space for further analysis. The gravity dual backgrounds can be found by performing a TsT (T-duality, 'shift' where the parameter of deformation enter, T-duality) transformation involving two of the angles that parametrize the $U(1)$ directions \cite{frolov4}, or by using the T-duality group\cite{ozer1}. An extended discussion for $\beta$-deformed Sasaki-Eistein dualities is presented in \cite{zaffaroni1}, where giant gravitons are also analyzed.

Therefore, it is  obvious that there is an extensive attempt to extend and understand better the $AdS/CFT$ correspondence using  theories other than $\cN=4$ supersymmetric Yang-Mills. In this paper we investigate  semi-classical string solutions in general $Y^{p,q}$ and $L^{p,q,r}$ manifolds. Work in this direction has been done for the very special case of $AdS_5\times T^{1,1}$ examined in \cite{kim1,pons1,schvellinger1}. Moreover,a study for the case of BPS massless geodesics and their dual long BPS operators has been done in \cite{benvenuti3} for $Y^{p,q}$ manifolds and in \cite{benvenuti4} for $L^{p,q,r}$. Dual giant gravitons have been studied in \cite{martelli2} and recently giant magnons and spiky strings moving in a sector of $AdS_5\times T^{1,1}$ have been examined in \cite{benvenuti5}.

In this paper we mainly work on the gravity side and examine the motion of the string along some  $U(1)$ directions in Sasaki-Einstein spaces
which is localized at $\r=0$ in the $AdS$ space. We will see that in some cases it is difficult to find acceptable string solutions, due to the constraints imposed on the Sasaki-Einstein parameters. For the solutions we find, we present the energy-spin relation. The energy depends on the manifold considered i.e. on $p,\,q,\,r$, but the momenta enter in the dispersion relation in a completely transcendental way.
We also present an extensive discussion on point-like BPS string solutions.
Notice that we do not examine the  string dynamics in $AdS_5$ which are identical to the maximally supersymmetric case, since all the equations can be decoupled.

\section{The backgrounds}
\subsection{$Y^{p,q}$ Metrics}
The Sasaki--Einstein metrics $Y^{p,q}$ on $S^2\times S^3$ can be presented  in the following local form \cite{gauntlett2}:
\ben
\label{se1}
  d s^2 &=& \frac{1-cy}{6}(d\theta^2+\sin^2\theta
      d\phi^2)+\frac{1}{w(y)q(y)}
      d y^2+\frac{q(y)}{9}(d \psi-\cos\theta  d \phi)^2 \nonumber\\
      & + &  {w(y)}\left[d \alpha +f(y) (d\psi-\cos\theta
      d \phi)\right]^2,
\een
or more compactly
\be
 d s^2 =d s^2(B)+w(y)[d \alpha+A]^2,
\ee
where
\ben
w(y) & = & \frac{2(a-y^2)}{1-cy}~ , \nn\\
q(y) & = & \frac{a-3y^2+2cy^3}{a-y^2}~ , \nn\\
f(y) & = & \frac{ac-2y+y^2c}{6(a-y^2)}~.
\een
For $c=0$ the metric takes the local form of the standard
homogeneous metric on $T^{1,1}$. Generally we can scale the constant $c$ to 1 by a diffeomorphism, and this is what we do in the rest of the paper.

To make the space $B$ a smooth complete compact manifold we should fix the coordinates appropriately \cite{gauntlett2}. The parameter $a$ is restricted to the range
\be
0<a<1~.
\ee
To make  the base
$B_4$ an axially squashed $S^2$ bundle
over the round $S^2$ one can choose the ranges of the
coordinates $(\theta,\phi,y,\psi)$ to be $0\le\theta\le \pi$,
$0\le \phi \le 2\pi$, $y_1\le y\le y_2$ and $0\le \psi \le 2\pi$. The parameter $\psi$ is the
azimuthal coordinate on the axially
squashed $S^2$ fibre and the round sphere $S^2$ parametrized by $(\theta,\phi)$.
Also, by choosing the above range for $a$, the following conditions
of $y$ are satisfied: $y^2<a$, $w(y)>0$ and  $q(y) \geq 0$. The equation $q(y)=0$ is cubic and has three real roots, one negative and two positive. Naming the negative root $y_{q-}$ and the smallest positive root $y_{q+}$ we must choose the range of the coordinate $y$ to be
\be\label{yqy}
y_{q-}\leq y\leq y_{q+}~,
\ee
with the boundaries corresponding to the south and north poles of the axially  squashed $S^2$ fibre.
Also, it is necessary to have $p/q$ rational in order to have  a complete manifold. Note that
\ben\label{yqyq}
y_{q+}-y_{q-}=\frac{3 q }{2 p}\equiv \xi~,
\een
where $\xi$ is defined for later use.
Therefore, if the roots $y_{q+},\,y_{q-}$ are rational  we speak of quasi-regular Sasaki-Einstein manifolds with the property that the volume of these manifolds having a rational relation to the volume of the $S^5$. However the rationality of $p/q$ can be achieved even in cases that the two roots are irrational which gives irregular Sasaki-Einstein metrics.

Using the expressions \eq{yprop} presented in the appendix A, we can express $y_{q-}$ in terms of $\xi$
\ben
y_{q-}=\frac{1}{2}(1-\xi-\sqrt{1-\frac{\xi^2}{3}})~,
\een
where $0<\xi<\sqrt{3}$. Since $y_{q-}$ is the root of the qubic,  $a$ can be expressed in terms of $\xi$
\ben\label{axi}
a=\frac{1}{18} \left(9-3 \sqrt{9-3 \xi^2}+4 \xi^2 \sqrt{9-3 \xi^2}\right)~,
\een
and in order to ensure that $y_{q+}$ is the smallest positive root we constrain $\xi$ to the range $0<\xi<3/2$. If we prefer, we can express $a$ in terms of $p,\,q$ using \eq{yqyq}
\be\label{apq}
a=\frac{1}{2}-\frac{p^2-3q^2}{4 p^3}\sqrt{4 p^2 -3 q^2}~,
\ee
then the period of $\a$ is given by $2 \pi l$ where
\ben\label{aper}
l=\frac{q}{3q^2-2 p^2+p(4p^2-3 q^2)^{1/2}}~.
\een

\subsection{$L^{p,q,r}$ Metrics}
The metric of this manifold is \cite{cvetic1}
\ben
 ds_{p,q,r}^2 &=& (d\xi+\sigma)^2 + ds_{[4]}^2~,
\label{metricL}
\een
where
\ben
ds_{[4]}^2 &=& \frac{\rho^2}{4\D(x)} dx^2 + \frac{\rho^2}{h(\theta)}d\theta^2 + \frac{\D(x)}{\rho^2}
\left(\frac{\sin^2\theta}{\alpha}d\phi+\frac{\cos^2\theta}{\beta}d\psi\right)^2 \\
 && \qquad+ \frac{h(\theta)\sin^2\theta\cos^2\theta}{\rho^2}\left(\frac{\a-x}{\alpha}d\phi-\frac{\b-x}{\beta}d\psi\right)^2
\een
and
\ben
\sigma &=& \frac{(\a-x)\sin^2\theta}{\a} d\phi + \frac{(\b-x)\cos^2\theta}{\b}d\psi~,\\
\D(x) &=& x(\a-x)(\b-x)-\mu ~,\\
\rho^2 &=& h(\theta)-x ~,\\
h(\theta) &=& \alpha\cos^2\theta + \beta \sin^2\theta~.
\een
Here $p,\, q$ and $r$ are relative positive coprime integers and $0<p\leq q,\,0 < r < p+q$ and $p,\,q$ are coprime to $s=p+q-r$. The metrics depends on two non-trivial parameters since $\a,\,\beta,\,\mu$ are constants, and we can set one of them equal to a non-zero number by rescaling the other two and $x$. The function $\D(x)$ plays a similar role to the function $f(y)$ in the $Y^{p,q}$ manifold, so $x$ should be restricted between the two lowest roots of $\D(x)=0$, namely $x_1$  and $x_2$, where
\be\label{xqx}
x_1<x<x_2~.
\ee
Moreover, in order to have a smooth geometry in $5$ dimensions the parameters should satisfy $\a,\b \geq x_2$ where $x_2\geq x_1\geq 0$, which imply the  already presented inequalities for $p,\,q,\,r$. The constants appearing in the metric are related to the roots of $\D(x)$ as follows:
\be
\mu= x_1 x_2 x_3,\,\,\,\,\quad
\a+\b=x_1+x_2 +x_3,\,\,\,\,\quad
\a\b=x_1 x_2+x_1x_3+x_2 x_3,
\ee
where $x_3$ is the other root of $\D(x)$.

The metrics \eq{se1} of $Y^{p,q}$ can be derived as a special case of $L^{p,q,r}$ when $p+q=2r$, which implies $\a=\b$.
The coordinate transformation is
\be\label{cc}
\bar{\xi}=-2 \a,\,\,\,\,\bar{\phi}=\frac{\psi +\phi}{2}+3 \a,\,\,\,\,\bar{\psi}=\frac{\psi-\phi}{2}+3\a,\,\,\,\,\bar{\theta}=\frac{\theta}{2},\,\,\,\,\bar{x}=\frac{(2 y +1)\bar{\a}}{3}~,
\ee
with $\mu$ related to $a$ by
\be
\bar{\mu}=\frac{4}{27}(1-a)\bar{\a}^3~ ,
\ee
where we redefine the coordinates and the constants of $L^{p,q,r}$ using bars, in order to distinguish them from the ones of $Y^{p,q}$.

We also present the metric of $AdS_5$ in the Hopf coordinate system although we will be using the time element only
\be
d s^2=-\cosh^2\r d t^2+d\rho^2 +\sinh^2\rho(d\beta_1^2+\cos^2 \b_1 d\b_2^2+\sin^2 \beta_1 d\beta_3^2)~,
\ee
where the ranges of the angles are $0\leq\b_1\leq\pi/2,\,0\leq\b_2,\,\b_3\leq2\pi$.

\section{String solutions in $Y^{p,q}$ background }
\subsection{Equations of motion and conserved quantities}
In this section we present some spinning string solutions  in the $Y^{p,q}$ manifold.
We will fix the angle $\theta$ and hence are allowing the string to move on a circle of the round sphere $S^2$ parametrized by the coordinate $ \phi$. On the squashed sphere the string can move on its azimuthal coordinate $\psi$, and sit at a constant value $y_0$ between the north and south poles. This value will be chosen carefully by solving the equations of motion. Finally, the string can move on the principle $S^1$ bundle over $B$ parametrized by $\a$. Notice that each of the directions that the string is allowed to spin has a $U(1)$ symmetry.
As usual the global time is expressed through the world-sheet time as $t=\k \t$, and the string is localized at the point $\rho=0$. The Polyakov action in the conformal gauge is given by
\ben\nonumber
S&=&-\frac{\sqrt\l}{4 \pi}\int d\t d\s\big[- (-\dot{t}^2+t'{}^2)+\frac{1-y}{6}(-\dot{\theta}^2+\theta'{}^2)+\frac{1}{w q}(-\dot{y}^2+y'{}^2) \\\nonumber
&&+(\frac{1-y}{6}s_\th^2+\frac{q}{9}c_\th^2+w f^2 c_\th^2)(-\dot{\phi}^2+\phi'{}^2)+(\frac{q}{9}+w f^2)(-\dot{\psi}^2+\psi'{}^2) +w(-\dot{\a}^2+\a'{}^2)\\\nonumber &&
-2 c_\th(\frac{q}{9}+w f^2)(-\dot{\psi}\dot{\phi}+\psi'{}\phi'{})+
2 w f(-\dot{\a}\dot{\psi}+ \a'{}\psi'{})-2 w f c_\th(-\dot{\a}\dot{\phi}+\a'{}\phi'{})\big]~.
\een
For convenience we do not write explicitly the dependence of $y$ in the functions $f,\,w$ and $q$.
The classical equations of motion for constant $\theta$ and $y$ take the form
\ben\nonumber
&&\frac{1-y}{6}(-\dot{\phi}^2+\phi'{}^2)s_{2\theta} + ( \frac{q}{9}+ w f^2)\big(s_{2\theta}(\dot{\phi}^2-\phi'{}^2)+2 s_\theta(-\dot{\psi}\dot{\phi}+\psi'{}\phi'{})\big)\\
&&\qquad\qquad\qquad\qquad\qquad\qquad\qquad\qquad\qquad\qquad+2 w f s_\theta(-\dot{\a}\dot{\phi}+\a'{}\phi'{})=0~, \label{eom1}\\\nonumber
&&\frac{s_\theta^2}{6}(\dot{\phi}^2-\phi'{}^2)+(\frac{Q}{9}+ A_1 )\big(c_\theta^2(-\dot{\phi}^2+\phi'{}^2)-\dot{\psi}^2+\psi'{}^2-2 c_\theta(-\dot{\psi}\dot{\phi}+\psi'{}\phi'{})\big)\\
&&\qquad\qquad\qquad+W(-\dot{\a}^2+\a'{}^2)+2 A_3(-\dot{\a}\dot{\psi}+ \a'{}\psi'{}-c_\theta(-\dot{\a}\dot{\phi}+a'{}\phi'{})\big)=0~,\label{eom2}
\een
\ben
&&\partial_{\b}[ \gamma^{\b\d}\big(w \partial_\d \a +w f ( \partial_\d\psi-c_\theta\partial_\d\phi)\big) ] =0~,\label{eom3}\\
&&\partial_\b[\gamma^{\b\d}\big(\frac{1-y}{6}s_\theta^2\partial_\d \phi+(\frac{q}{9}+w f^2)(c_\theta^2 \partial_\d \phi-c_\theta\partial_\d\psi)
- w f c_\theta \partial_\d a\big)]=0~,\label{eom4}\\
&&\partial_\b[\gamma^{\b\d}\big((\frac{q}{9}+w f^2)(\partial_\d \psi-c_\theta\partial_\d \phi)+ w f \partial_\d \a\big)]=0~,\label{eom5}
\een
where we have used the conventions
\be
A_1\equiv\partial_y(w f^2),\quad A_2\equiv\partial_y((w q)^{-1}),\quad A_3\equiv\partial_y(w f),\quad Q\equiv\partial_y q,\quad  W\equiv\partial_y{w}
\ee
and we have written the three last equations in a more compact form since they are satisfied trivially for our ansatze.
In addition to the above equations, we get two more which come from the variation of the action with respect to the worldsheet metric. These are equivalent to the components of the energy-momentum tensor being set to zero. The Virasoro constraints read
\ben\nonumber
&&\frac{1-y}{6}s_\theta^2\dot{\phi}\phi'{}+(\frac{q}{9}+ w f^2 )\big(c_\theta^2\dot{\phi}\phi'{}+\dot{\psi}\psi'{}-c_\theta(\dot{\phi}\psi'{}+\dot{\psi}\phi'{})\big)
+w \dot{\a}\a'{}\\
&&\qquad\qquad\qquad\qquad\qquad\qquad\qquad\quad+w f \big(\dot{\a}\psi'{}+\dot{\psi}\a'{}-(\dot{\a}\phi'{}+\dot{\phi}\a'{})c_\theta\big)=0,\label{vc1}\\\nonumber
&&-\k^2+\frac{1-y}{6}s_\theta^2(\dot{\phi}^2+ \phi'{}^2)+ w (\dot{\a}^2+ \a'{}^2)+2 w f \big(\dot{\a}\dot{\psi}+\a'{}\psi'{}-(\dot{\a}\dot{\phi}+\a'{}\phi'{})c_\theta\big)
\\\nonumber
&&\qquad\qquad\qquad\quad\quad+
(\frac{q}{9}+ w f^2 )\big(c_\theta^2(\dot{\phi}^2+ \phi'{}^2)+\dot{\psi}^2+ \psi'{}^2
-2 c_\theta(\dot{\phi}\dot{\psi}+\phi'{}\psi'{})\big)
\\
&&\qquad\qquad\qquad\qquad\qquad\qquad\qquad\,\,\,\,\,\,+\frac{1-y}{6}(\dot{\theta}^2+\theta'{}^2)+\frac{1}{w q} (\dot{y}^2+y'{}^2)=0 ,
\label{vc2}
\een
where only in the last equation \eq{vc2} we include the terms corresponding to a non-constant $\theta$ and $y$ for later use in section 3.2.

The symmetry of $Y^{p,q}$ admits three conserved charges which are the angular momenta corresponding to strings rotating along the $\a,\,\phi$ and $\psi$ directions. Moreover, there exists one more conserved quantity, the classical energy, which is generated by the translational invariance along $t$. All of them are presented below:
\ben\label{1e}
E&=&\frac{\sqrt{\l}}{2 \pi}\int_{0}^{2 \pi}d \sigma\, \k ~,\\\label{1ja}
J_a&=&\frac{\sqrt{\l}}{2 \pi}\int_{0}^{2 \pi}d \sigma\,(w \dot{a}-  w f c_\theta \dot{\phi} +w f \dot{\psi})~,\\\label{1jphi}
J_\phi&=&\frac{\sqrt{\l}}{2 \pi}\int_{0}^{2 \pi}d \sigma\,\big( -w f c_\theta \dot{a}+(\frac{1-y}{6}s_\theta^2  + \frac{q}{9}c_\theta^2 +w f^2 c_\theta^2)\dot{\phi}-(\frac{q}{9} +w f^2)c_\theta \dot{\psi}\big),\\\label{1jpsi}
J_{\psi}&=&\frac{\sqrt{\l}}{2 \pi}\int_{0}^{2 \pi}d \sigma\,\big( w f \dot{a}-(\frac{q}{9} +w f^2)c_\theta\dot{\phi}+(\frac{q}{9} +w f^2)\dot{\psi}\big)~.
\een
Let us also define  the new quantities $J_{tot}=J_\a+J_\phi+J_\psi$, $\cE=  E /\sqrt{\l}$, $\cJ_i = J_i /\sqrt{\l}$ and  $\cJ_{tot} = J_{tot} /\sqrt{\l}$ for later use. The conserved quantity corresponding to the total $SU(2)$ angular momentum is
\be
J^2=J_\theta^2+\frac{1}{s_\theta^2}(J_\phi+c_\theta J_\psi)^2+J_\psi^2~.
\ee
Finally, for point-like stings localized at constant points on $\theta$ and $y$, the following identity holds
\be
\sqrt{\l}\,\k^2=\dot{\a} J_\a+ \dot{\phi }J_\phi+ \dot{\psi } J_\psi~.
\ee
In the following we will choose an ansatz where the string is moving in $Y^{p,q}$ along the three angles $\a,\,\phi,\,\psi$ and is at rest along all the other directions
\ben\label{ansatz1}
&&\a=\o_1\t+m_1 \s,\qquad\phi=\o_2\t+m_2 \s,\qquad\psi=\o_3\t+m_3 \s~,\\
&&\theta=\theta_0\,\,\,\,\mbox{and}\,\,\,\, y=y_0,
\een
where $\theta_0,\,y_0$ are constants and their exact values should be chosen to be consistent with the solutions of the equations of motion and the Virasoro constraints. Notice also, that due to the periodicity condition in the global coordinates of the manifold on $\sigma$, the winding numbers have to be integers.
For the linear dependence on $\t,\,\s$ of \eq{ansatz1}, the equations of motion \eq{eom3}, \eq{eom4} and \eq{eom5}  for $\a,\,\phi$ and $\psi$ respectively,  are trivially satisfied.

\subsection{Discussion on BPS solutions}

In this section we discuss the BPS point-like solutions. The R-symmetry in the field theory is dual to the canonically defined Reeb Killing vector field $K$ on the Sasaki-Einstein manifolds  \cite{martelli1,benvenuti1}, given by
\be
K=3 \frac{\partial}{\partial\psi}-\frac{1}{2}\frac{\partial}{\partial\a}
\ee
and the R-charge is equal to
\be\label{Rc}
Q_R=2 J_\psi-\frac{1}{3}J_\a~.
\ee
Now in order to express the Hamiltonian in terms of the momenta, we are initially considering  a general situation, where all the parameters in the internal manifold are dependent on $\t$ :
\be
\theta=\theta(\t), \qquad y=y(\t),\qquad \a=\a(\t),\qquad \phi=\phi(\t),\qquad \psi=\psi(\t),
\ee
and later we will focus on the string configurations mentioned in the previous section. The reason we are doing this, is  to show how the general BPS solutions behave if we generalize our ansatz and activate simultaneously the motion on all angles. The process is equivalent to finding massless geodesics and is examined in \cite{benvenuti3,martelli2}.

We start by  expressing the energy in terms of the momenta. Now since we consider motion on the $\theta$ and $y$ coordinates, we have the-non zero conjugate momenta
\ben
J_y=\frac{\sqrt{\l}}{2 \pi}\int_{0}^{2 \pi}d \sigma\,\frac{1}{w\, q}\dot{y}~,\qquad
J_\theta=\frac{\sqrt{\l}}{2 \pi}\int_{0}^{2 \pi}d \sigma\,\frac{1-y}{6} \dot{\theta}~ .
\een
It is straightforward to substitute in the second Virasoro constraint the velocities in terms of their momenta and get
\ben\label{spins}
\k^2=\frac{1}{\l}\Big(w\, q\, J_y^2 +\frac{6}{1-y}(J^2-J_\psi^2)+\frac{1}{w} J_\a^2+\frac{9}{q}(J_\psi-f J_\a)^2\Big) ~.
\een
The energy of the string is given by \eq{1e}, and is equal to the conformal dimension $\D$ of the dual operator, and to find the lower bound of it, we should express \eq{spins} in terms of the R-charge. Using \eq{Rc}, 
we obtain via the algebra in \eq{spins}:
\ben
\D^2=\big(\frac{3}{2}Q_R\big)^2+\frac{1}{w q}(J_a+3 y Q_R)^2+ w q J_y^2 + \frac{6}{1-y}(J^2-J_\psi^2)~.
\een
Since $y_{q+}<1$, which is the upper bound of $y$, and $J^2\geq J_\psi^2$, all the terms in the above equation are positive, which leads to the inequality $\D \geq 3/2\, Q_R$. The solutions generated by the equality correspond to BPS operators, and in order to saturate the bound, all the following equations  must be satisfied
\be\label{bps}
J_y=0,\qquad J_\theta=0,\qquad J_\phi=-c_\theta J_\psi,\qquad J_\a=-3 y Q_R .
\ee
The two first equations fix $\theta$ and $y$ to unknown constants. The next two can be used to determine the relationship between $\a,\,\phi,\,\psi$ using the constants  $y,\,\theta$. The situation is now similar to our initial configuration in the previous section where we considered $y,\,\theta$ as constant, with the difference that the $\t$ dependence of the $U(1)$ angles is now unknown, and needs to be determined by the equations of motion and the Virasoro constraints. In order to find BPS solutions we need to solve the equations of motion
(24-28) together with \eq{bps} and use \eq{vc2} to calculate the energy. The first Virasoro constraint \eq{vc1} is trivially satisfied.

It is more convenient to proceed by solving first the equations \eq{eom3}, \eq{eom4} and \eq{eom5} for constant $y$ and $\theta$, since their solutions are very special. As we said above, these equations are satisfied trivially for angles with  linear dependence on $\tau$, i.e. the one written in  \eq{ansatz1} where for the point-like case is equivalent to set all the $m_i$ zero.  The rest of the solutions we get constrain $y$ to live on its maximum or minimum values, which means on the poles of the squashed sphere. More specifically the solutions are
\ben\label{1a1}
&&\a=\o_1\t,\qquad \phi=\o_2\t,\qquad\psi=\o_3\t  \quad\,\,\,\mbox{or}\\
&& y=y_{q\pm},\qquad \phi=\o_2\t,\qquad \ddot{\a}=\frac{1-y}{6y}\ddot{\psi}. \label{2a2}
\een
However, when we take account of the boundary conditions in \eq{2a2}, we see that $\dot{\psi}=0$, since we are at the poles of the squashed sphere. This fixes $\psi$ to be constant and the other angles to have  linear dependence on $\t$. 
So the only way to satisfy the above equations for the non-constant angles is to have a linear dependence on $\tau$, where the $\theta$ and $y$ are not yet fixed to a specific value.

Bearing the above results in mind, we proceed by finding general solutions that satisfy the two first equations of motion \eq{eom1}, \eq{eom2}, 
together with the BPS equations \eq{bps} and present some solutions, starting with
a solution which was also found in \cite{benvenuti3}
\ben\label{benv}
\dot{\phi} = 0,\qquad \dot{\a} =-\frac{\dot{\psi}}{6}.
\een
We see that this solution is valid for any $y$ that satisfies the inequality \eq{yqy}, since in this case, the last equation of \eq{bps} is satisfied trivially. Additionally, the third equation  in \eq{bps} is satisfied also trivially and hence $\theta$ can take any constant value inside the region where it is defined. However notice that in  order for the solution to satisfy 
(26-28), the $\a$ and $\psi$ angles must have a linear dependence on $\tau$, and hence
\be
\a=-\frac{\o_3}{6}\t,\qquad\psi=\o_3\t.
\ee
The energy for this solution is equal to $E=\sqrt{\l}|\dot{\psi}|/3$ and the conserved momenta in this case are
\be\label{bps1}
J_\a=-\frac{2}{3}\sqrt{\l}  y \dot{\psi},\qquad J_\phi=\frac{1}{9}\sqrt{\l}(y-1) c_\theta \dot{\psi}, \qquad J_\psi= -\frac{1}{9}\sqrt{\l} (y-1)\dot{\psi}
\ee
and are related each other by
\be\label{bps2}
J_\phi=- c_\theta\, J_\psi=\frac{1-y}{6y} c_\theta\,J_\a~.
\ee
In this case the energy can be written as
\ben\label{bpsen1}
E= \frac{3}{|(y-1)(c_\theta-1)-6y|}\, |J_{tot}|~.
\een
Notice that in the above relation the factor of proportionality is independent of $a$, and hence on the manifold considered.

Another set of solutions are:
\ben\label{bps1b}
\theta=0,\qquad y=1\pm\frac{\sqrt{1-a}}{\sqrt{3}},\qquad\dot{\a}=\frac{\dot{\phi}-\dot{\psi}}{6} ,
\een
where again $\a, \phi, \psi$ have to be linear with $\tau$ in order for the above expressions to satisfy \eq{eom3}, \eq{eom4} and \eq{eom5}. The $y$ solution is a special case of \eq{nonacc} for $\theta=0$ and as we show in the next section, it is unfortunately always greater than $y_{q+}$ except in the limit $y=y_{q+}=1$. This is the case where $a=1$ and we are not going to examine it any further. However, to be more accurate here, we have to set $\dot{\phi}=0$, since $\theta=0$ on the pole of $S^2$ and there is no meaning in defining rotation along $\phi$ direction. Hence, the corresponding equation in \eq{bps1b} should be modified to $\dot{\a}=-\dot{\psi}/6$.

More interesting are the following solutions where $y$ lives on its boundaries
\ben\label{bps3}
y=y_{q\pm},\qquad \theta=0 ,\qquad\dot{\phi}=0,\qquad
\dot{\a}=-\frac{y+2}{6 y} \dot{\psi},
\een
and the dependencies of the non-constant angles on $\t$ are linear.
This solution is acceptable since the inequality \eq{yqy} is satisfied. However, by considering the boundary conditions, the solution become trivial, since we are on the poles of the squashed sphere
where  rotation along the $\psi$ direction cannot be defined. Hence, we have to impose $\dot{\psi}=0$ and then the whole string ansatz becomes static.

One  could  possibly find, other non-interesting solutions at the limits  $a=0,\,1$, but it seems that there are no other BPS solutions than the ones presented above, which allow $a$ to be at other points except zero and one.
The solution \eq{bps3}  has as its main property to restrict $y$ to the boundaries of \eq{yqy} and to localise the string on the two three-submanifolds obtained by the initial manifold for $y=y_{q\pm}$, and denoted by $\S_-\cong S^3/\cZ_{p+q}$,\,$\S_+\cong S^3/\cZ_{p+q}$. The cones over these Lens spaces are divisors of $C(Y^{p,q})$ and hence supersymmetric submanifolds \cite{martelli1}. This is because the induced volume form on $\S$ is equal to the four-form $\cJ\equiv J\wedge J/2$ , where $J$ here is the K\"{a}hler symplectic form. Hence, these cones are calibrated with respect to $\cJ$.


In the following sections, non-BPS point-like, as well as extended string solutions, will be examined.

\subsection{One angle solution}

In this section we examine the simplest case, where only the angle which parametrizes some $U(1)$ direction of the manifold is turned on. In this case it is known that classical spinning string solutions that wrap around the circle do not exist due to diffeomorphism invariance, or equivalently because the first Virasoro constraint forces the metric element in the spinning direction to vanish, or make the string ansatz trivial. In $Y^{p,q}$ manifolds  however, we just mention that the diagonal metric elements in these three Killing vector directions can not vanish for $y$ in the range \eq{yqy}.

For the $g_{\a\a}$, this is obvious since it is equal to zero for $y=\pm\sqrt{a}$ and we already know that these values do not satisfy \eq{yqy}.
\begin{figure}
\centerline{\includegraphics[scale=0.44]{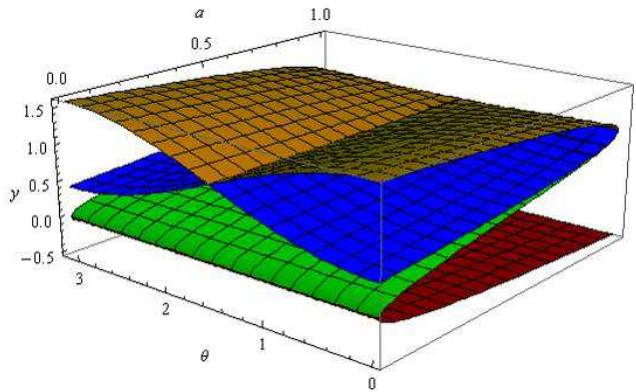}}
\caption{We plot $y_{\pm}=1\pm \sqrt{1-a}c_\theta^2/\sqrt{3}$ together with the $y_{q_+},\,y_{q-}$ versus $a,\,\theta$. The  $y_{q_-},\,y_{q+}$ surfaces are colored red and green respectively, while $y_{-},\,\,y_{+}$ colored blue and orange respectively. From the figure we notice that both $y_{\pm}$ are greater than $y_{q+}$ for the whole range of $a$ and $\theta$. }
\end{figure}
For the $g_{\phi\phi}$ element, the situation is more complicated since it is equal to zero for
\be\label{nonacc}
y_{\pm}=1\pm\frac{ \sqrt{1-a}c_\theta^2}{\sqrt{3}}~ .
\ee
Obviously the $y_{+}$ solution is discarded since is bigger than one, but also the $y_{-}$ solution is outside the desirable area as can be seen in Figure 1.
The last  diagonal metric element $g_{\psi\psi}$ is zero for $y_{\pm}=1\pm \sqrt{1-a}/\sqrt{3}$, where $y_-$ is the lower bound of the previous solution and obviously this metric element cannot be zero too.

Let us begin by looking for point-like string solutions. By allowing the string to move only along the $\a$ direction, and using the ansatz $\a= \o_1 t$, the only non-trivially satisfied equation that we need to solve is \eq{eom2}, which takes the simple form:
\be
\left(2-\frac{2 (a-1)}{(y-1)^2}\right) \o_1^2 =0~,
\ee
and has solutions
\be\label{y1}
y_{\pm}=1 \pm \sqrt{1- a}~.
\ee
Comparing these solutions with the $y_{q\pm}$ (Figure 2), we see that  $y_-$ is an acceptable solution, since $y_{q-}\leq y_-\leq y_{q+}$ for the whole range of $a$.
\begin{figure}
\includegraphics[width=70mm]{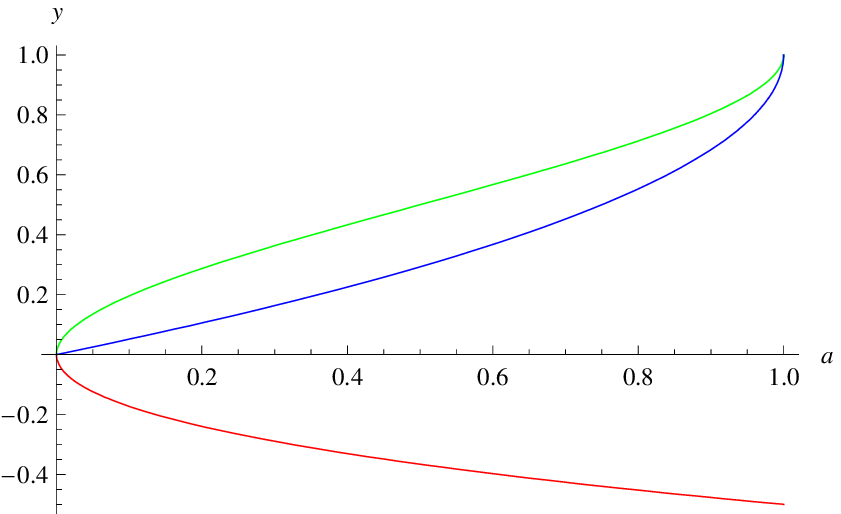}
\includegraphics[width=70mm]{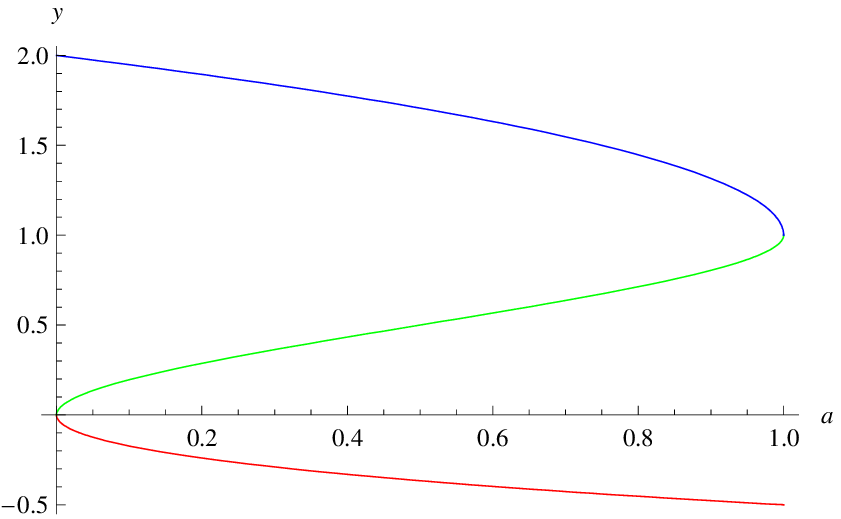}
\caption{Plotting the roots $y$ (blue) of \eq{y1} with $y_{q-}$ (red), $y_{q+}$ (green) versus $a$. This color mapping to the solution will be the same in the other plots too. In the first plot we see that the smaller solution $y_-$ is between the two roots of  $q(y)=0$ and in the second one that the greater root $y_+$ is outside the allowed area since $y_{+}>1$. }
\end{figure}
Then, for $y=y_-$ the second Virasoro constraint \eq{vc2} gives
\be\label{oneang1}
\k^2=w\, \o_1^2\Rightarrow \k^2=4 \left(1-\sqrt{1-a}\right) \o_1^2~.
\ee
Then using \eq{1e} - \eq{1jpsi}, we get for energy and the non zero conserved charges
\ben
E&=& \sqrt{\l}\, \sqrt{w\, \o_1^2}~,\\
J_\a&=& \sqrt{\l}\, w\, \o_1~.
\een
Combining the above relations we end up with
\ben
E= \sqrt{\frac{1}{w}\,J_\a^2}\Rightarrow\quad
E=  \frac{\sqrt{\l}}{2}\, \sqrt{\frac{1}{1 - \sqrt{1 - a}}\cJ_\a^2}\, .  \label{e1}
\een
Note that the energy depends linearly on $\cJ_\a=\cJ_{tot}$ and depends on the manifold $Y^{p,q}$, since $a$ is related to $p,\,q$ by the equation \eq{axi}, with the factor of proportionality being a monotonically decreasing function with respect to $a$.
On the other hand, the function of the energy in terms of $a,\,\o_1$ is a monotonically increasing function with respect to $a$ as can be seen from \eq{oneang1}. As a final remark, we mention that all the calculations are independent of the angle $\theta$, which can be chosen to be any constant angle.


For completeness we now can consider a static string along the $\a$ direction with $\a= m_1 \sigma$. The solution of the equation of motion remains the same, given by \eq{y1}. The only difference is that all the momenta are now equal to zero and the energy is expressed as
\be\label{oneang22}
\k^2=w m_1^2\Rightarrow E= 2\sqrt{\l}\, \sqrt{\left(1-\sqrt{1-a}\right) m_1^2}~.
\ee

Following a similar procedure we choose a different angle $\phi$ with a similar ansatz and consider a point-like string motion with $\phi=\o_2 t$. 
Then, the non-trivial equations are \eq{eom1} and \eq{eom2}, which take the form
\ben
&&\frac{(a-1)s_{2\theta}}{18 (y-1)}\o_2^2=0~,\\
&&\frac{ a-7-6 (y-2) y+(a-1)c_{2\theta}}{36 (y-1)^2}\o_2^2=0~,
\een
and do not give any new real solution, since the first equation is solved for $\theta=n \pi/2,\,n=(0,1,2)$ and the second one can not be solved for real $y$. The situation is similar when we choose the static string ansatz $\phi=m_2 \s$, where the above equations remain the same with $\o_2$ replaced by $m_2$, and obviously again they do not give any new solutions.

Finally, for the third angle which parametrizes the remaining $U(1)$ direction, again consider the string ansatz $\psi=\o_3 t$.
The only non-trivial equation of motion is \eq{eom2}
\be\label{eooom}
\frac{a-4-3 (y-2) y}{18 (y-1)^2}\o_3^2 =0~,
\ee
which has solutions $y= 1\pm \sqrt{-1+a}/\sqrt{3}\,$ and are not real. The situation is similar for the ansatz $\psi=m_3 \s$, since the equation \eq{eooom} remains the same, with only difference the replacement of $\o_3$ by $m_3$.

Summarizing, if we restrict the string to rotate or wrap only along one $U(1)$ direction in $Y^{p,q}$ as we have done above, there is only one possible configuration. It is the point-like string moving along the fibre direction $\a= \o_1 \t$. This is one spin solution, where the energy  is proportional to $\cJ_\a$ and depends on the manifold $Y^{p,q}$ in a way that the factor of proportionality is a decreasing function of $a$.
We expect that the solution \eq{y1} is not BPS. More specifically the Hamiltonian here is
\be
H=\sqrt{\l}\frac{1}{2} w \o_1^2=2 \sqrt{\l} \left(1-\sqrt{1-a}\right) \o_1^2~.
\ee
Hence the R-charge and the conformal dimension of the dual operator can be written as
\be\nonumber
Q_R=-\frac{\sqrt{\l}}{3} w \o_1,\,\,\,\D^2=(\frac{3}{2} Q_R)^2+\l (4 a-w)\o_1^2=(\frac{3}{2} Q_R)^2+ \frac{\sqrt{1-a}}{4 (1- \sqrt{1-a})}J_{\a}^2~.
\ee
It is obvious that  $\D>3/2\, Q_R$ always, since the expression $(4 a-w)$ is always positive, and become equal to zero only in the limits $a=0,\,1$. Hence, the solution is non-BPS.

\subsection{The two Angle Solutions}

Here we examine some solutions where the strings are moving along two $U(1)$ directions.
We are looking for possible string solutions, with motion along the directions $\a, \psi$, motivated by the fact that motion in these directions gives  one BPS solution. Considering point-like strings and the simple ansatz\footnote{Whenever we are writing an ansatz we suppose that all the parameters $\o_i,\,m_i$ are non-zero.}
\be
\a = \o_1 \t,\qquad \psi = \o_3 \t,
\ee
we get only one non-trivial equation \eq{eom2}. This is solved by
\ben\label{two1}
y_{\pm}=1\pm\frac{\sqrt{(1-a) (2 \o_1-\o_3) (6 \o_1+\o_3)}}{\sqrt{3} (2 \o_1-\o_3)},\,\,\,\mbox{or}\,\,\,\o_1=-\frac{\o_3}{6}.
\een
We do not examine solutions in the limits $a=0,\,1$ and in the rest of the analysis we will ignore solutions here. The solution $\o_1=-\o_3/6$ is BPS and is examined in a previous section. The other solution for $y$, is acceptable in a region which will be specified. Notice, that for
\be
\o_1=-\frac{y+2}{6 y} \o_3~,
\ee
which is the solution derived in \eq{bps3}, the solution \eq{two1} for $y$ becomes BPS and lives on $y=y_{q\pm}$, but by considering the boundary conditions becomes static.

For the general case, first of all, one must find the values for which the square root is real and these are for
\ben
&&\o_3>0,\,\,\,\o_1 > \frac{\o_3}{2}\,\,\,\mbox{or}\,\,\,\o_1<-\frac{\o_3}{6}\\
&&\o_3<0,\,\,\,\o_1 <\frac{\o_3}{2}\,\,\,\mbox{or}\,\,\,\o_1>-\frac{\o_3}{6}.
\een
Then we can see  that indeed there exists solutions of $y$  that are between $y_{q-}$ and $y_{q+}$. Actually, the two solutions of \eq{two1} are equivalent depending on the sign of numbers $\o_1,\,\o_3$, i.e. if they are positive or negative. What one can say directly, is that the solution for $y$ with RHS equal to 'one minus a positive quantity', is the one that could be acceptable in some intervals. In order keep the presentation simpler we choose $\o_1,\,\o_3>0$.  To give a visual picture of how the solutions behave we plot the surfaces for some random relation between $\o_1, \o_3$, say $\o_1=n \,\o_3$ in Figure 3.
\newline
\newline
We see that $y_{-}$ is an acceptable solution for
\be\label{conda}
a> \frac{4 (7+18 n)}{(1+6 n)^3}~,
\ee
which also implies $n>1/2$ in order for $a$ to be smaller than $1$.

\begin{figure}
\includegraphics[width=65mm]{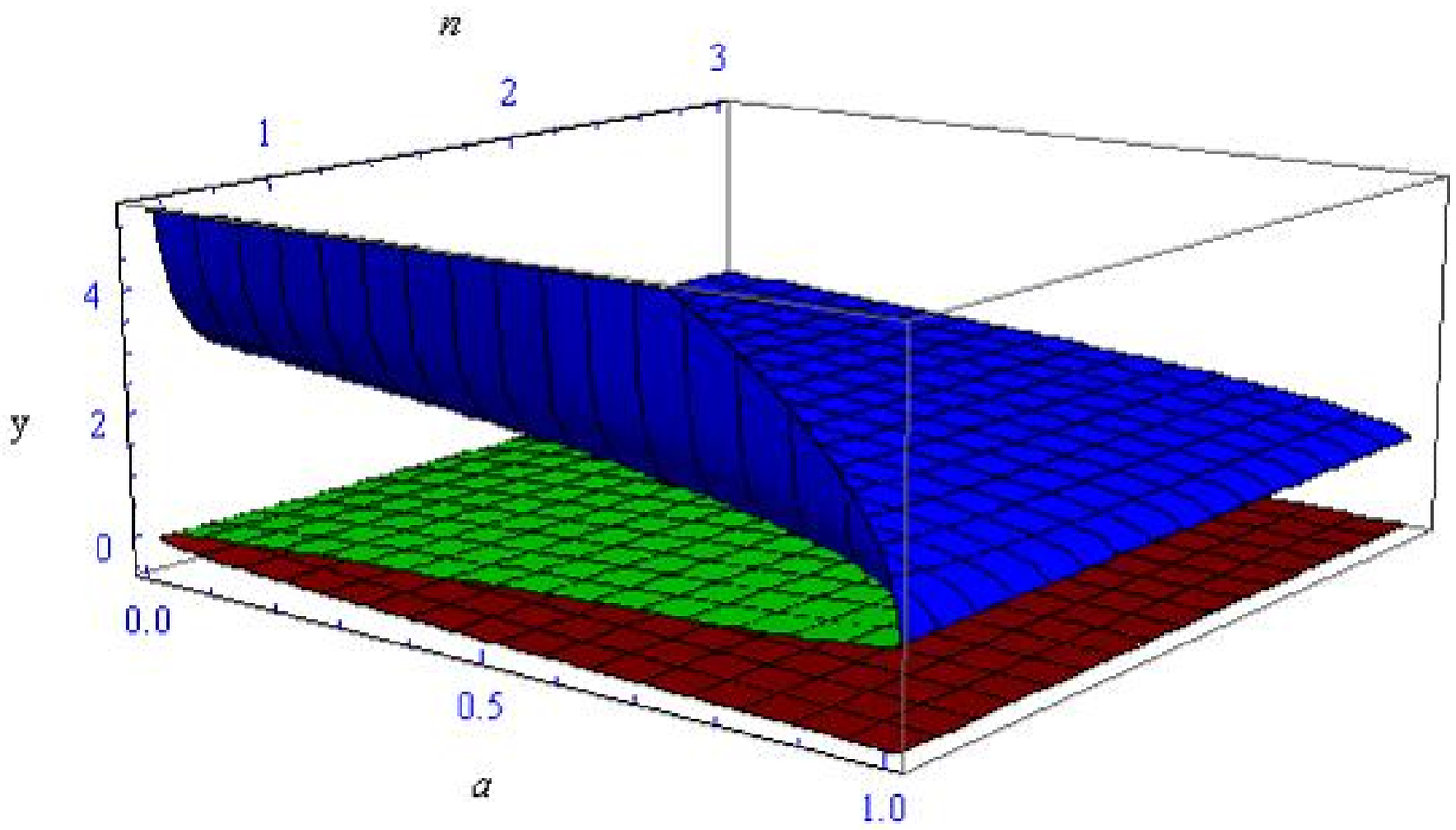}\qquad
\includegraphics[width=65mm]{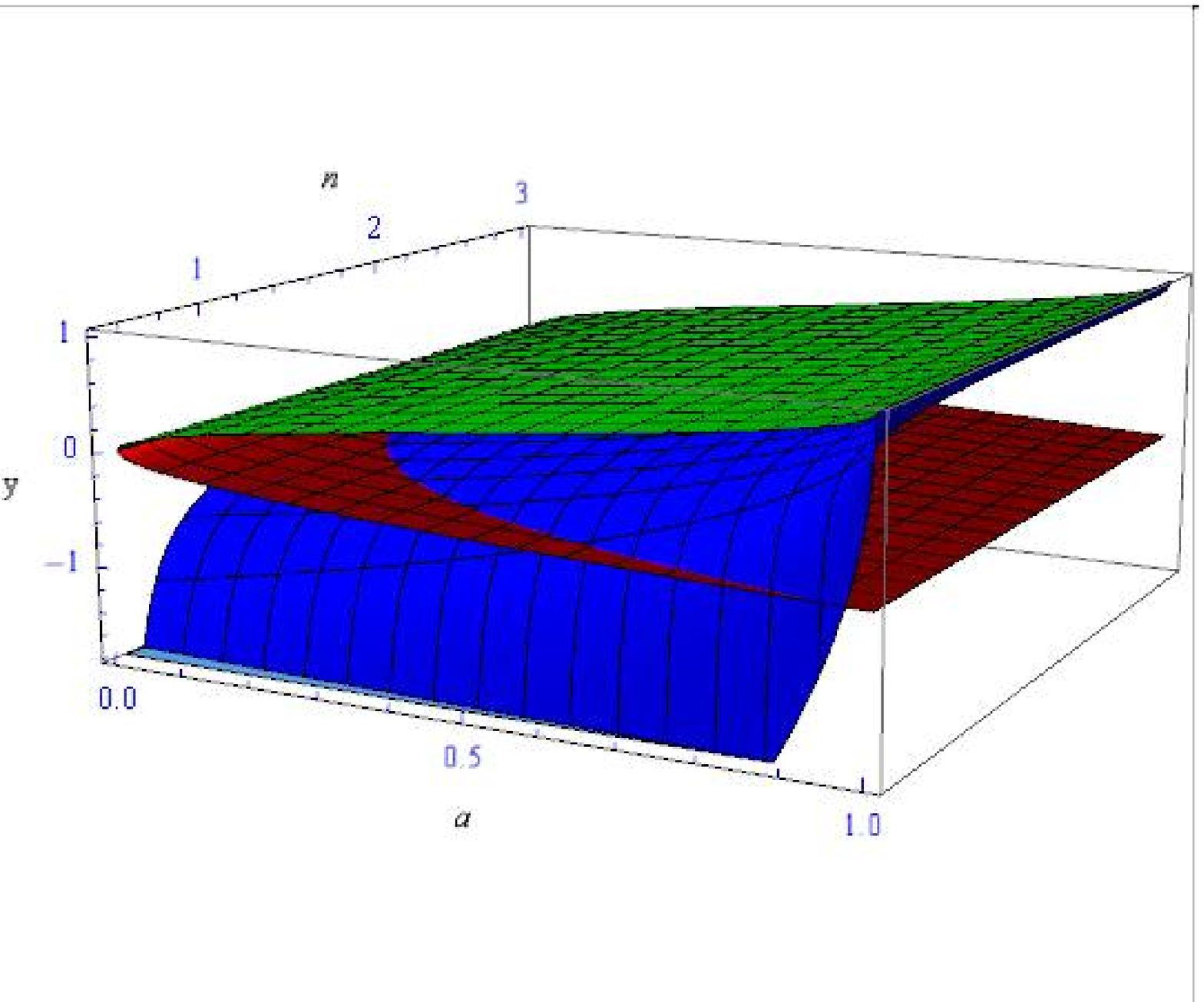}
\caption{We plot $y_{\pm}$ versus $n, a$ where $\o_1=n\, \o_3$ with $n>1/2$. In the first plot with $y_{+}$ we clearly see that there is no acceptable solution. In the second plot, the $y_{-}$ give acceptable solutions.}
\end{figure}
For the general case, the equation \eq{vc2} gives
\be\label{en2}
\k^2=4 \o_1^2-\frac{(6  \o_1 +\o_3)\sqrt{(1-a) (2 \o_1-\o_3) (6 \o_1+\o_3)}}{3 \sqrt{3}} ~,
\ee
where we have substituted the solution $y$.
\newline
\newline
The conserved charges for our solution, given from \eq{1ja}, \eq{1jphi}, \eq{1jpsi} are
\ben
J_\a&=&\sqrt{\l}\Big(4 \o_1+\frac{4 (-3 \o_1+\o_3)}{3} \sqrt{\frac{(1-a)  (6 \o_1+\o_3)}{3 (2 \o_1-\o_3)}}\Big)~,\\
J_\phi&=&-\sqrt{\l} \frac{\o_3 c_\theta}{3}  \sqrt{\frac{(1-a) (6 \o_1+\o_3) }{3 (2 \o_1-\o_3)}} ~,\\
J_\psi&=&\sqrt{\l}\frac{\o_3}{3} \sqrt{\frac{(1-a) (6 \o_1+\o_3) }{3 (2 \o_1-\o_3)}}~,
\een
where \eq{two1} used. Notice the relation $J_\phi=-c_\theta J_\psi$. It seems that the energy depends on the momenta in a transcendental way. This is due to the complicated expression of $y$ which depends on $\o_{1,3}$, and enters in the momenta through the functions $w,\,q,\,f$. On the other hand,  as we can see from \eq{en2}, the energy is a monotonically increasing function with respect to $a$ when $\o_1,\,\o_3$ are fixed. This behavior is similar to the one angle solution we found before, but it will be more interesting to have an energy-spin relation.

To finish, let us consider the example  of $Y^{2,1}$. To find the corresponding value of $a$ we can use \eq{axi} and find that
\ben\label{y21}
a&=&\frac{1}{4} \left(2-\frac{\sqrt{13}}{8}\right)\simeq 0.387327 ~.
\een
Substituting in \eq{conda}, we see that  indeed we can get solutions for $y$ that are inside the desirable interval when $\o_1>\,0.8568\,\o_3$, or equivalently $n > 0.8568$.
To simplify things even more we can choose $\theta=0$, and solve for $\o_1,\,\o_3$ in terms of $J_\a,\,J_\psi$ to get the energy-spin relation. The energy-spin relation looks lengthy and complicated and seems to be completely transcendental, so we choose not to present it here.


Generalizing  the ansatz by adding a $\sigma$ dependence on the angle $\a$  and allowing the string to spin along this direction, we search for string solutions using
\be
\a = \o_1 \t+ m_1 \sigma,\qquad \psi = \o_3 \t.
\ee
The non-trivial equations that needs to be solved in this case are \eq{eom2} and \eq{vc1}. One acceptable solution is
\ben\label{a1}
\o_1=\frac{\o_3}{6},\qquad y=a, \qquad  a=\frac{2 \o_3^2}{9 m_1^2+\o_3^2}~.
\een
By considering $a<1$, and choosing $\o_1,\,\o_3>0$, we get  $|m_1|>\o_3/3$. To satisfy the condition \eq{yqy}, we have to restrict  $m_1$ further  as
\be\label{bound22}
|m_1|>\frac{\o_3}{\sqrt{3}}\Rightarrow a<\frac{1}{2}~.
\ee
Hence solutions that satisfy the above condition are acceptable as can be also seen clearly in Figure 4.
\begin{figure}\label{two2t}
\centerline{\includegraphics[width=69mm]{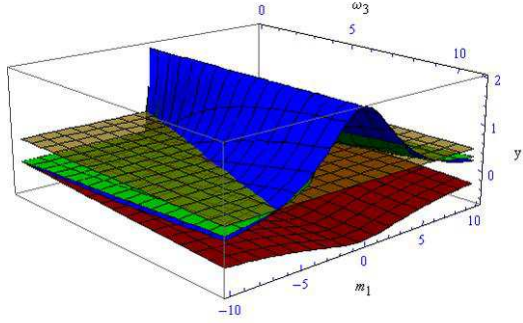}}
\caption{Plotting  $y,\, y_{q+},\,y_{q-}$  versus $\o_3$, $m_1$. The transparent plane is at $a=1/2$. When \eq{bound22} satisfied, $a<1/2$ and hence our solution lives in the area between the $y_{q+},\,y_{q-}$. Also notice that for $a>1$  there  is no green surface, since the values of $y_{q+}$ become complex, as expected.}
\end{figure}
\newline
\newline
To calculate the energy we use \eq{vc2} which gives
\be\label{twoe}
\k^2=\frac{(15 m_1^2 -\o_3^2)\o_3^2}{3 \left(9 m_1^2+\o_3^2\right)}~,
\ee
and the corresponding charges are given by
\be\label{twomom2}
J_\a=0, \qquad J_\phi=\sqrt{\l}\frac{ c_\theta \left(\o_3^2-3 m_1^2\right)}{3 \left(9 m_1^2+\o_3^2\right)}\o_3, \qquad J_\psi=\sqrt{\l}  \left(-\frac{1}{3}+\frac{4 m_1^2}{9 m_1^2+\o_3^2}\right)\o_3.
\ee
Before we continue, we point out that one has to check if the function of $a$ we get from \eq{a1} can give rational values to the expression $y_{q+}-y_{q-}$. This was not done in the previous cases since $a$ was not dependent on $\o$ or $m$, and hence it can take any appropriate value between zero and one, which would make the subtraction of the two roots of $q(y)=0$ rational. In this case, one way to solve this problem is to equate the expression for $a$, given in \eq{axi}, with our solution \eq{a1}, and see if we can get rational solutions in the acceptable interval for $\xi$. One more reason to express $\o_3$ in terms of $a$ is that is more preferable to choose a manifold $Y^{p,q}$, and from that to specify the allowed values for $\o_i,\,m_i$, instead of proceeding in the reverse direction. Using \eq{axi} and \eq{a1}, we get for $\o_3, m_3$
\ben\label{dd}
\o_3=\pm\frac{3 \sqrt{a} }{\sqrt{2 - a}}m_1=
\pm \frac{3 \sqrt{9+\sqrt{9-3 \xi^2} \left(-3+4 \xi^2\right)}}{\sqrt{27+\left(3-4\xi^2\right) \sqrt{9-3 \xi^2}}} m_1 \equiv\, \pm D m_1~,
\een
where  $D$ is defined from the above equation. We are going examine the solutions with the plus sign which imply $m_1>0$.
For $0<\xi<3/2$, the multiplicative factor satisfies $0<D<3$, and by imposing the constraint \eq{bound22}, we are limited in the interval
\be
\xi<\frac{\sqrt{3}}{2}\Leftrightarrow a<\frac{1}{2}~ ,
\ee
which also means that $D<\sqrt{3}$. The corresponding conserved charges in terms of $a,\,m_1$ are
\be\label{twomom}
J_\a=0, \qquad J_\phi=-\sqrt{\l}\frac{(1-2 a)\sqrt{a} c_\theta }{3 \sqrt{2- a}}m_1, \qquad J_\psi=\sqrt{\l}\frac{(1-2 a) \sqrt{a} }{3 \sqrt{2-a}}m_1~,
\ee
or using $\o$'s
\be\label{twomommmmm}
J_\phi=-\sqrt{\l}\frac{(1-2 a) c_\theta }{9} \o_3, \qquad J_\psi=\sqrt{\l}\frac{(1-2 a) }{9} \o_3~.
\ee
The sum of the momenta is
\be
J_{tot}=\frac{\sqrt{\a} (2 \a-1) (\cos{\theta}-1)}{3 \sqrt{2-\a}}m_1 ~.
\ee
Finally, the relation between the charges is
\be
J_\phi=-c_\theta J_\psi=-\frac{c_\theta}{1-c_\theta} J_{tot}~.
\ee
Notice, that the zeroth momentum $J_\a$ can follow from a more general solution such as $y=a$ and $\o_1=\o_3/6$.
\newline
To find the energy, we use \eq{vc2} to get
\be
\k^2=\frac{a (5-4 a)}{2-a} m_1^2 ~,
\ee
which can be written in terms of $\o_3$ by solving \eq{dd} for $m_1$. As before for a fixed $\o_3$, the energy is a monotonically increasing function with respect to $a$ for $a<1/2$.
Using \eq{twomom}, we get the energy of our solution in terms of the momenta
\be\label{twoene1}
E=\sqrt{\l} \sqrt{\frac{a (5-4 a) }{2-a} m_1^2}= \frac{3\sqrt{5-4 a}}{(1-2 a)(1-c_\theta)} J_{tot} ~.
\ee
It is also interesting to insert the $\xi$ parameters in the expressions we calculate using \eq{dd}.
Then the energy in terms of the momenta and the rational number $\xi$ is
\be\label{twoene2}
E=\frac{3 \sqrt{3}}{(3-4 \xi^2)(1-c_\theta)} \sqrt{\frac{27+2(3-4 \xi^2 )  \sqrt{9-3 \xi^2}}{3-\xi^2}} J_{tot}~.
\ee

From \eq{twoene1} or \eq{twoene2} we see that the energy is proportional to the momenta and depends on $a$. The exact dependence of the energy multiplied by $(1-c_\theta)$, in order to avoid the dependence on the third parameter $\theta$ and to be able to plot a surface, is presented in Figure 5.
\begin{figure}\label{twoplot}
\includegraphics[width=60mm]{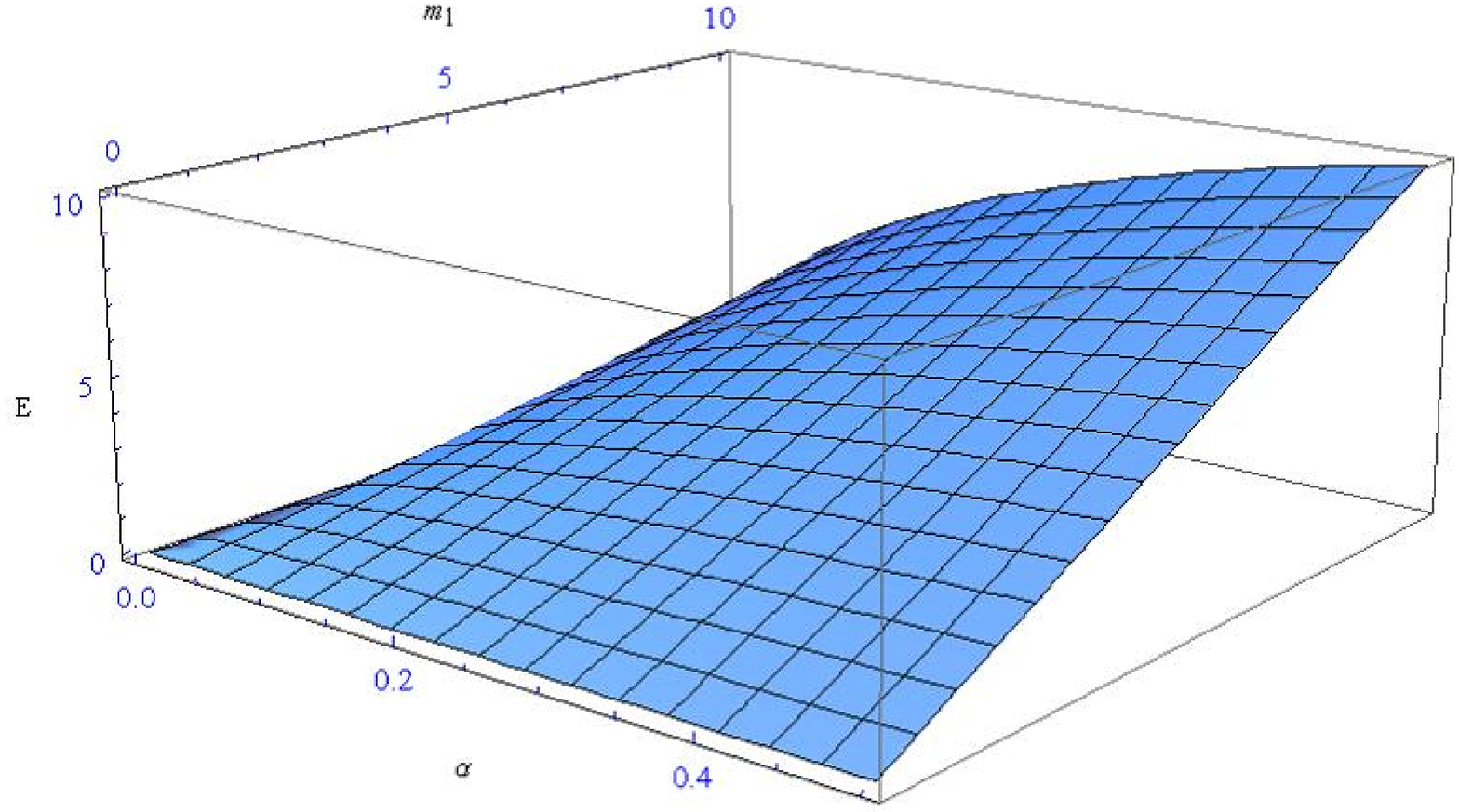}\qquad
\includegraphics[width=65mm]{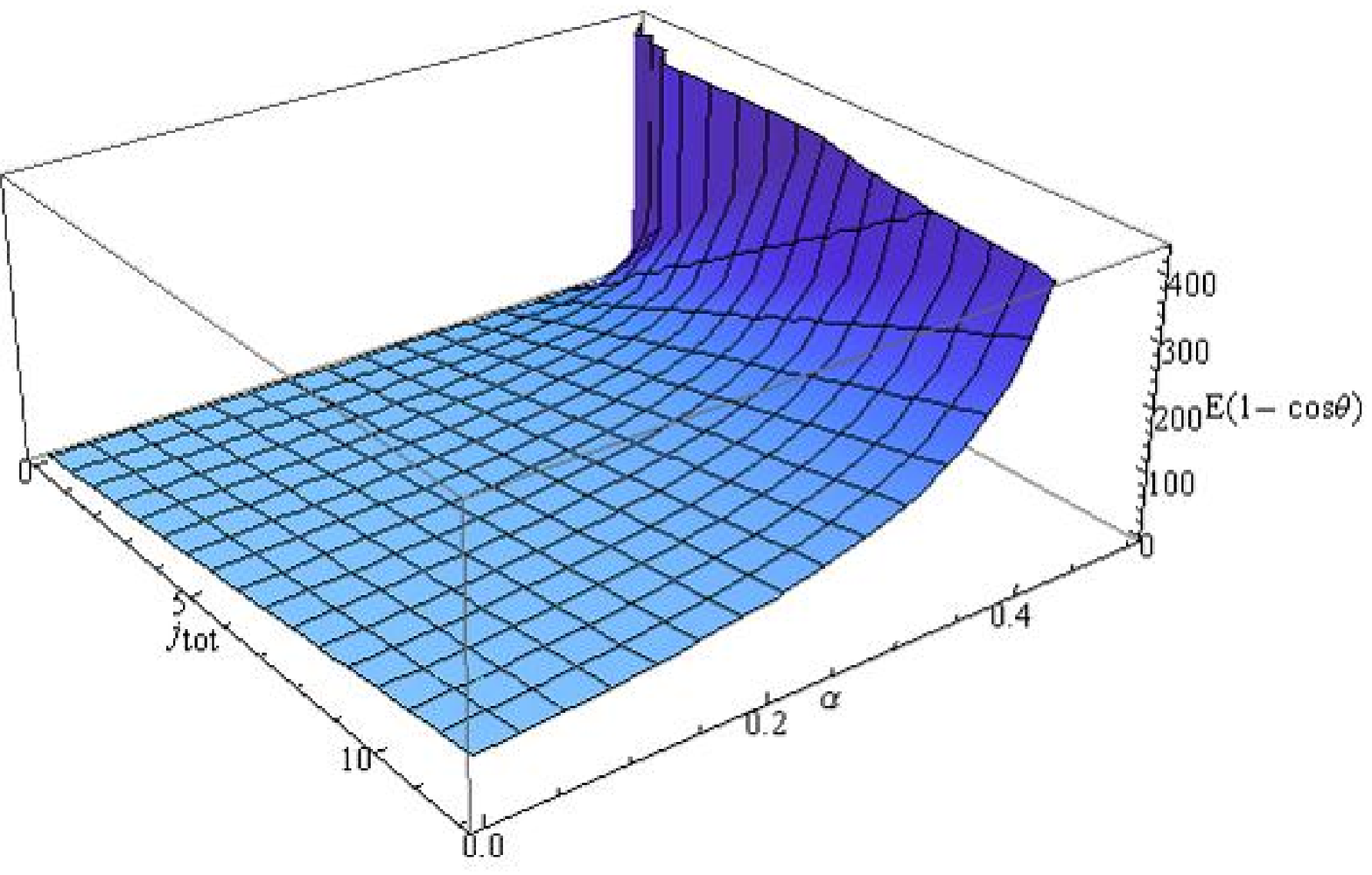}
\caption{In the first plot is the energy versus $m_1$, $a$.  In the second plot the expression $E \cdot (1-c_\theta)$ is plotted  versus $J_{tot}$, $a$, and one can notice the sharp rise of the energy as $a\rightarrow 1/2$. The plot range on energy is restricted intentionally to finite region in order to have a clear shape of the surface.}
\end{figure}
We point out here that the function of energy in terms of the momenta again has a linear dependence where the factor of proportionality depends on the manifold $a$ and is monotonically increasing with respect to $a$. However, the form of this factor does not seem to follow a specific pattern between the different string solutions. This occurs mainly because the parameter $y$ depends on the $\o,\, m$ in a different way for each string solution. Also notice that the momenta are not taking continuous values but are quantized on $a$ and $m_1$. Hence, the energy will not be a continuous function of the parameters, even if we plot it as continuous in order to show its behavior.

To finish the analysis for this case we mention that there are other solutions of \eq{eom2} and \eq{vc1}, but these solutions are excluded since there is no $a$ that satisfies  the rationality constraint of $y_{q+}-y_{q-}$ and at the same time gives integer winding numbers.

\section{String solutions in $L^{p,q,r}$ background }

\subsection{Equations of motion and conserved quantities}

In this section we construct solutions for strings  moving on the $L^{p,q,r}$ manifold. The configuration is chosen to be similar with the analysis of the $Y^{p,q}$ manifold. Firstly, we choose the string  to sit at a constant angle $\theta$. In this manifold, the role of the previous $y$ coordinate is played by the coordinate $x$, and so we restrict the string to be localized at a constant point $x_0$, which has to respect the constraint \eq{xqx}. For some configurations, the points $\theta$ and $x$ that the string is sitting can be chosen arbitrary, but in most cases  the equations of the system constrain at least one of them. As for the string dynamics,  we are going to consider motion along some of the three $U(1)$ directions and try to find  the energy-spin relation and how the energy is related on the properties of the general manifold.

Furthermore, we are not going to analyze the string dynamics in the $AdS_5$ since these are identical to the maximally supersymmetric case. To simplify things we are localizing the string also at $\rho=0$ on $AdS_5$ and expressing the global time through the world-sheet time by $t=\k \t$.
Thus, the Polyakov action in the conformal gauge is given by
\ben\nonumber
S&=&-\frac{\sqrt\l}{4 \pi}\int d\t d\s\big[- (-\dot{t}^2+t'{}^2)+ (-\dot{\xi}^2+\xi'{}^2)+\frac{\r^2}{4 \D}(-\dot{x}^2+x'{}^2)+
\\\nonumber
&&+\frac{\r^2}{h}(-\dot{\theta}^2+\theta'{}^2)+\Big(\frac{(\a-x)^2}{\a^2}s_\theta^2+\frac{\D s_\theta^2+h c_\theta^2(\a-x)^2}{\r^2 \a^2}\Big)s_\theta^2(-\dot{\phi}^2+\phi'{}^2)
+ \\\nonumber
&&+\Big(\frac{(\b-x)^2}{\b^2}c_\theta^2+\frac{\D c_\theta^2+h s_\theta^2(\b-x)^2}{\r^2 \b^2}\Big)c_\theta^2 (-\dot{\psi}^2+\psi'{}^2)+
\\\nonumber
&&+2\Big((\a-x)(\b-x)+\frac{\D-h(\a-x)(\b-x)}{\r^2}\Big)\frac{c_\theta^2 s_\theta^2}{\a \b}
(-\dot{\psi}\dot{\phi}+\psi'{}\phi'{})
+\\\nonumber
&&+2 \frac{\a-x}{\a}s_\theta^2(-\dot{\xi}\dot{\phi}+\xi'{}\phi'{})+ 2 \frac{\b-x}{\b}c_\theta^2(-\dot{\xi}\dot{\psi}+\xi'{}\psi'{})~.
\een
Again, we do not write the dependence of the functions on their arguments, in order to simplify the presentation.
Also notice that $\a$ should not be confused with the same letter used in the case of $Y^{p,q}$ manifolds to name the angle.
Before we start writing down the equations of motion, we define some new quantities, in order to simplify the presentation. Hence we identify the three expressions which are multiplied by $(-\dot{\phi}^2+\phi'{}^2),\, (-\dot{\psi}^2+\psi'{}^2)$ and $(-\dot{\psi}\dot{\phi}+\psi'{}\phi'{})$, in the above action, which also correspond to metric elements, with the functions $d_1(x,\theta)\equiv g_{\phi\phi},\,d_2(x,\theta)\equiv g_{\psi\psi}$ and $d_3(x,\theta)\equiv g_{\phi\psi}$ respectively. The partial derivatives of these functions are presented in Appendix A, and will be used later.
\newline
\newline
The equations of motion for the $\theta,\,x$  are:
\ben\nonumber
&&
\partial_\theta d_1(-\dot{\phi}^2+\phi'{}^2)+\partial_\theta d_2 (-\dot{\psi}^2+\psi'{}^2)+2 \partial_\theta d_3 (-\dot{\psi}\dot{\phi}+\psi'{}\phi'{})\\
&&\qquad\qquad\qquad\quad+2\frac{a-x}{a}s_{2\theta}(-\dot{\xi}\dot{\phi}+\xi'{}\phi'{})- 2 \frac{\b-x}{\b}s_{2\theta}(-\dot{\xi}\dot{\psi}+\xi'{}\psi'{}) =0~,\\\nonumber
&&\partial_x d_1(-\dot{\phi}^2+\phi'{}^2)+\partial_x d_2 (-\dot{\psi}^2+\psi'{}^2)+2\partial_x d_3(-\dot{\psi}\dot{\phi}+\psi'{}\phi'{})\\
&&\qquad\qquad\qquad\qquad\qquad\qquad-\frac{2}{\a}s_\theta^2(-\dot{\xi}\dot{\phi}+\xi'{}\phi'{})- \frac{2}{\b}c_\theta^2(-\dot{\xi}\dot{\psi}+\xi'{}\psi'{})=0~,
\een
where we have used the fact that the string is localized at two fixed points $\theta_0,\,x_0.$
The equations of motion for the three $U(1)$ directions $\phi,\,\psi,$ and $\xi$ are
\ben\label{eoml3}
&&\partial_\k\big(\gamma^{\k\l}(d_1\partial_\l\phi+d_3\partial_\l\psi+2 \frac{\a-x}{\a}s_\theta^2\partial_\l\xi)\big)=0~ ,\\\label{eoml4}
&&\partial_\k\big(\gamma^{\k\l}(d_2\partial_\l\psi+d_3\partial_\l\phi+2 \frac{\b-x}{\b}c_\theta^2\partial_\l\xi)\big)=0~ ,\\\label{eoml5}
&&\partial_\k\big(\gamma^{\k\l}(2 \frac{\a-x}{\a}s_\theta^2\partial_\l\phi+2 \frac{\b-x}{\b}c_\theta^2\partial_\l\psi +\partial_\l\xi)\big)=0~.
\een
Note that they are trivially satisfied for the linear ansatz we choose.
\newline
\newline
Furthermore, the Virasoro constraints are given by
\ben\nonumber
&&\dot{\xi} \xi'+d_1 \dot{\phi}\phi'{}
+d_2\dot{\psi}\psi'{}+ d_3 (\dot{\psi}\phi{}'+\psi{}'\dot{\phi})+\frac{\a-x}{\a}s_\theta^2(\dot{\xi}\phi'{}+\xi'{}\dot{\phi}) \\ &&\qquad\qquad\qquad\qquad\qquad\qquad\qquad\qquad\qquad\,\,\,+\frac{\b-x}{\b}c_\theta^2(\dot{\xi}\psi'{}+\xi'{}\dot{\psi})=0~,\\\nonumber
&&\k^2= \dot{\xi}^2+\xi'{}^2+ d_1(\dot{\phi}^2+\phi'{}^2)
+d_2 (\dot{\psi}^2+\psi'{}^2)+2 d_3(\dot{\psi}\dot{\phi}+\psi'{}\phi'{})\\
&&\qquad \qquad \qquad \qquad \qquad
+2 \frac{\a-x}{\a}s_\theta^2(\dot{\xi}\dot{\phi}+\xi'{}\phi'{})+ 2 \frac{\b-x}{\b}c_\theta^2(\dot{\xi}\dot{\psi}+\xi'{}\psi'{})~.
\een
Finally, the conserved charges associated to the three $U(1)$ isometries are
\ben\label{1jal}
J_\xi&=&\frac{\sqrt{\l}}{2 \pi}\int_{0}^{2 \pi}d \sigma\,(\dot{\xi}+\frac{\a-x}{\a}s_\theta^2\dot{\phi}+\frac{\b-x}{\b}c_\theta^2\dot{\psi}) ~,\\\label{1jphil}
J_\phi&=&\frac{\sqrt{\l}}{2 \pi}\int_{0}^{2 \pi}d \sigma\,\big( \frac{\a-x}{\a}s_\theta^2\dot{\xi} +d_1 \dot{\phi}  + d_3 \dot{\psi} \big) ~,\\\label{1jpsil}
J_{\psi}&=&\frac{\sqrt{\l}}{2 \pi}\int_{0}^{2 \pi}d \sigma\,\big( \frac{\b-x}{\b}c_\theta^2\dot{\xi}+ d_3 \dot{\phi}+d_2 \dot{\psi}   \big)~,
\een
where the classical energy is given by \eq{1e}.
In the next section we use these equations to find some string solutions on $L^{p,q,r}$.
\newline
\newline
The ansatz  for the string motion we consider have a linear dependence with $\tau$ and $\sigma$ and are the following
\ben
&&\xi=\o_1\t+m_1 \s,\qquad \phi=\o_2\t+m_2 \s, \qquad \psi=\o_3\t+m_3 \s\\
&&\theta=\theta_0\,\,\,\,\mbox{and}\,\,\,\, x=x_0,
\een
where $\theta_0,\,x_0$  are constants and we are also setting $\mu=1$.

\subsection{One angle solution}

It is straight-forward to see that in the coordinate system we chose for the metric,  there are solutions for point-like strings moving on the direction $\xi$, since the metric element $g_{\xi\xi}$ is constant. As we also commented in a previous section, a general property of the Virasoro constraints is that they do not allow any extended spinning string solutions, where the strings are moving only along one dimension.

We start by considering the trivial case of $\xi=\o_1 \t$, where all the equations of motion and the first Virasoro constraint are satisfied trivially. The conserved momenta are
\ben
J_\xi=\sqrt{\l}\o_1,\,\,\,\,\,
J_\phi=\sqrt{\l}\frac{\a-x}{\a}s_\theta^2\o_1,\,\,\,\,J_{\psi}=\sqrt{\l}\frac{\b-x}{\b}c_\theta^2\o_1.
\een
The second Virasoro constraint gives the energy
\be\label{enl}
E = \sqrt{\l}\sqrt{\o_1^2} = |J_\xi|=\frac{2 \a\b}{4 \a\b-(\a+\b)x-(\a-\b)x~ c_{2\theta} }|J_{tot}|~,
\ee
which is presented in terms of the momenta. We see that the energy is proportional to the total spin and the factor of proportionality depends on $\a,\b$, and hence on the manifold $L^{p,q,r}$ considered.

Now consider a point-like string rotating along $\phi$ direction with $\phi=\o_2 \t$. The first two equations of motion give
\ben
\partial_\theta d_1(x,\theta)=0~,\qquad
\partial_x d_1(x,\theta)=0~~.
\een
In  the region that $\a,\,\b$ are defined, the only  real solution is when $\theta=0$, which makes the $g_{\phi\phi}$ element zero.
As a final possibility, consider the string moving according to  $\psi=\o_3 \t$, which has to satisfy the equations
\ben
\partial_\theta d_2(x,\theta)=0~,\qquad
\partial_x d_2(x,\theta)=0~.
\een
These have a real solution only for $\theta=\pi/2$ which makes the metric element in the direction of rotation equal to zero.

For completeness we mention that the corresponding static string ansatze $m_i\s$, gives an acceptable solution only for  a wrapping  around the $\a$ direction, and in this case all the conserved monenta are zero.

\subsection{The two angle solutions}

In this section we try to sketch a way of finding solutions of strings moving in two Killing vector directions simultaneously.

We choose to activate the angles $\xi,\,\psi$, and initially look for point-like string solutions
\be
\xi=\o_1 \t,\qquad\psi=\o_3 \t.
\ee
The equations of motion reduce to
\be
\partial_\theta d_2\, \o_3-2\frac{\b-x}{\b}s_{2\theta}\, \o_1=0~,\qquad
\partial_x d_2 \,\o_3-2\frac{c_{\theta}^2}{\b} \o_1=0~,
\ee
where  in order to check easier the inequality \eq{xqx}, and solve the above equations, it is convenient to give an appropriate value to angle $\theta$. By choosing  $\theta=\pi/4$ we get a solution
\ben\label{ltwo1}
\a=\b\pm\frac{3 \o_3}{\sqrt{2} \sqrt{-\b \o_3 (2 \o_1+\o_3)}},\qquad x=\frac{\a+ 5  \b}{6}~.
\een
To make sure the solution is real we need to constrain $\o_3$ by
\be
\o_1 < 0\quad\mbox{and}\quad 0 < \o_3 < -2\o_1 \quad\mbox{or}\quad  \o_1 > 0\quad\mbox{and}\quad -2 \o_1 < \o_3 < 0~.
\ee
Moreover, using the inequalities between $x_2,\,\a,$ and $\b$ we can easily see that  $x_2\leq (\a+ 5  \b)/6$, which means that  $x$ can only be equal to $x_2$. This seems to be a general feature for these solutions, since even for $\theta=\pi/3$, the situation is similar. But for a general angle $\theta$, it is more difficult to solve the equations of motion and identify this behavior. However, in the cases mentioned above, the solution $x$ is always  strictly greater than $x_2$, and hence not acceptable (Figure 6). There are also other solutions that could give acceptable answers, but need more extensive analysis.
\begin{figure}
\centerline{\includegraphics[width=70mm]{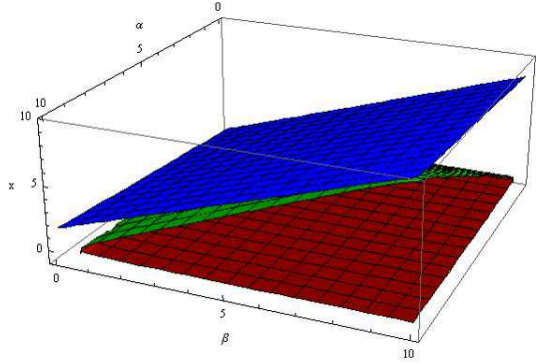}}
\caption{In the plot are presented the solutions of $\D(x)$, $x_1$ and $x_2$, together with $x$ \eq{ltwo1} versus the manifold parameters $\a,\b$. The solution $x$ is plotted with blue and we see that there is no region such that $x$ is between of $x_1,\,x_2$ where the manifold constraints are satisfied.}
\end{figure}

To obtain an extended string configuration consider the ansatz
\be
\a=\o_1 \t +m_1 \s,\,\,\,\,\,\,\psi=\o_3\t+m_3\s ,
\ee
which gives the following system of equations
\ben
&&\partial_\theta d_2 (-\o_3^2+m_3^2)-2\frac{\b-x}{\b}s_{2\theta} (-\o_1\o_3+m_1 m_3)=0~,\\
&&\partial_x d_2 (-\o_3^2+m_3^2)-\frac{2}{\b}c_{\theta}^2 (-\o_1\o_3+m_1 m_3)=0~,\\
&&\o_1 m_1+ d_2 \o_3 m_3 +\frac{\b-x}{\b}c_\theta^2(\o_1 m_3+m_1 \o_3)=0~.
\een
For $\o_1=m_1$ and $\o_3=m_3$ reduces to a single equation
\be
1+d_2 \frac{\o_3^2}{\o_1^2}+2 \frac{\b-x}{\b}c_\theta^2 \frac{\o_3}{\o_1}=0
\ee
and the simpler solution is for $\theta=\pi/4$,
\be\label{ert}
\o_1=-\frac{\o_3}{2},\qquad x=\frac{-2 + \a \b^2 + \b^3}{2 \b^2}~ .
\ee
In order this solution to satisfy the inequalities between $\a, \b,$ and $x$, is essential for the following inequality $|\a-\b|<2/\b^2$ to be satisfied. However, one can see that the solution is not acceptable because it is always greater than $x_2$. It is also interesting to examine the more complicated solutions, and see if they can satisfy \eq{xqx}.

Since it is complicated to check analytically if the solutions which involve a general angle $\theta$ satisfy the manifold constraints, we choose not to present them here, because our intention in this section was only to sketch the way of finding string solutions moving in two $U(1)$ directions.

\section{Conclusion}

In this paper, we initially examine the string motion on $Y^{p,q}$ along the $U(1)$ directions. To accept solutions of the equations of motion and the Virasoro constraints, we must make sure that they also satisfy  the Sasaki-Einstein constraints in a way we described above. Due to the presence of all these constraints, the number of acceptable string solutions is limited drastically. Then we show that when  the energy is expressed in terms of the conserved momenta, the factors multiplied with them depend on the manifold, i.e. on the parameter $a$, where they are monotonic functions with respect to it. Hence, the extrema of these functions occur for maximum or minimum values of $a$.
Moreover, the dispersion relation we found implies that the momenta enter into this relation in a transcendental way.

Furthermore, by looking at massless geodesics, we find that there is a unique BPS solution, which was already known. What we see here is that the string coordinates must depend linearly with time. Moreover, for this solution, the string can sit anywhere in the allowed $y$ interval satisfying \eq{yqy}.
An other point-like string solution was found which lives on the two  supersymmetric three-submanifolds $ S^3/\cZ_{p+q}$,\,$ S^3/\cZ_{p+q}$, obtained by the initial manifold for $y=y_{q\pm}$. However, by considering the boundary conditions, the solution becomes static, and we argue that there are no other point-like BPS solutions in the analysis presented above.

One can work similarly  in the  cohomogeneity two manifolds $L^{p,q,r}$, finding some point-like and classical string solutions. Again, we expect that when the energy is expressed in terms of the conserved momenta, it does not take the same form uniformly over the family of manifolds $L^{p,q,r}$, and basically the discussion we presented for the $Y^{p,q}$ manifold remain similar with the $L^{p,q,r}$ manifold. However, in this case, due to the large number of parameters, it is more difficult to check analytically in full generality  whether or not the solutions satisfy the manifold constraints. One can certainly try to find some more solutions for these manifolds.

An interesting extension of this work could be to consider strings having $\sigma$ dependence on one of the $y,\,\theta$ angles. By activating  simultaneously one more $U(1)$ angle for the string's motion, the analysis should not be difficult. However, simultaneous string motion on more directions, could lead to systems of differential equations that might be   difficult to  solve. Also, worth looking at, is the effect of the $\beta$ deformations on the string solutions on these manifolds. One can initially work with the point-like BPS solutions presented above, and try to derive the  $\beta$ deformed 'BPS condition'. It can then be checked if the BPS massless geodesics found above, remain undeformed. It must be possible to support these results accordingly in the dual beta deformed theory. Another interesting topic is to examine the string motion simultaneously in several $U(1)$ directions, and to analyse the energy-spin relations. Finally, it would be very good to identify the solutions found in this paper with the corresponding operators in field theory. Some of the above issues will be examined in a forthcoming publication \cite{giataganas3}.

\startappendix
\Appendix{Some formulas of Sasaki-Einstein manifolds}

The three roots of qubic satisfy
\ben\label{yprop}
y_{q+}+y_{q-}+y_3=3/2,\quad
y_{q+}y_{q-}+y_{q+}y_3+y_{q-}y_3=0,\quad
2 y_{q+}y_{q-}y_3=-a
\een
and also can be expressed in terms of $p,\,q$
\ben
y_{q\pm}=\frac{1}{4p}(2 p \pm 3 q-\sqrt{4 p^2-3 q^2}),\quad
y_3=\frac{1}{4p}(2 p +2\sqrt{4 p^2-3 q^2})
\een
and the period of $a$ \eq{aper}, can be rewritten in a more compact form
\be
l=-\frac{q}{4 p^2 y_{q+} y_{q-}} ,
\ee
which is always positive since $y_{q-}$ is negative. The volume $Y^{p,q}$ is given by
\ben
Vol(Y^{p,q})=\frac{q(2 p +\sqrt{4 p^2-3 q^2})l \pi^3}{3 p^2}
\een
and is bounded by
\be
Vol(T^{1,1}/\mbox{\cZ}_p)>Vol(Y^{p,q})>Vol(S^5/\mbox{\cZ}_2 \times \mbox{\cZ}_p).
\ee

\Appendix{Definition of functions used in $L^{p,q,r}$ string solutions}

In order to make the presentation shorter we define the following functions as
\ben
&&d_1(x,\theta)\equiv\Big(\frac{(\a-x)^2}{\a^2}s_\theta^2+\frac{\D s_\theta^2+h c_\theta^2(\a-x)^2}{\r^2 \a^2}\Big)s_\theta^2~,\\
&&d_2(x,\theta)\equiv\Big(\frac{(\b-x)^2}{\b^2}c_\theta^2+\frac{\D c_\theta^2+h s_\theta^2(\b-x)^2}{\r^2 \b^2}\Big)c_\theta^2 ~,\\
&&d_3(x,\theta)\equiv\Big((\a-x)(\b-x)+\frac{\D-h(\a-x)(\b-x)}{\r^2}\Big)\frac{c_\theta^2 s_\theta^2}{\a \b}~,
\een
which have the following partial derivatives. With respect to $\theta$:
\ben\nonumber
&&\partial_\theta d_1(x,\theta)=\frac{s_{2\theta}}{\a^2 (\a-\b)}\Big(\mu+\a (\a-\b) (\a-x) -\frac{4 \mu (\a-x)^2 }{ (\a+\b-2 x+(\a-\b) c_{2 \theta})^2}\Big)~,\\\nonumber
&&\partial_\theta d_2(x,\theta)=\frac{s_{2\theta}}{\b^2 (\a-\b)}\Big( \mu+\b (\b-\a) (\b-x)  -\frac{4 \mu (\b-x)^2 }{ (\a+\b-2 x+(\a-\b) c_{2 \theta})^2}\Big)~, \\\nonumber
&&\partial_\theta d_3(x,\theta)=\frac{ \mu s_{2 \theta} \left((x -\a c_{\theta}^2)c_{\theta}^2-(x - \b s_{ \theta}^2)s_{ \theta}^2\right)}{\a \b \left(-x+\a c_{ \theta}^2+\b s_{ \theta}^2\right)^2}
\een
and with respect to $x$:
\ben\nonumber
&&\partial_x d_1(x,\theta)=-\frac{s_{\theta}^2}{8 \a^2 \left(-x+\a c_{ \theta}^2+b s_{ \theta}^2\right)^2}\Big(4 \mu+\a \left(3 \a^2+2 \a \b+3 \b^2-8 (\a+\b) x+8 x^2\right)+\\
&&\qquad\qquad\qquad+4 (-\mu+\a (\a-\b) (\a+\b-2 x)) c_{2 \theta}+\a (\a-\b)^2 c_{4 \theta}\Big)~,\\ \nonumber
&&\partial_x d_2(x,\theta)=-\frac{c_{\theta}^2}{8 \b^2 \left(-x+\a c_{ \theta}^2+\b s_{ \theta}^2\right)^2}\Big(4 \mu+\b \left(3 \a^2+2 \a \b+3 \b^2-8 (\a+\b) x+8 x^2\right)+\\
&&\qquad\qquad\qquad+4 (\mu+(\a-\b) \b (\a+\b-2 x)) c_{2 \theta}+(\a-\b)^2 \b c_{4 \theta}\Big)~, \\\nonumber
&&\partial_x d_3(x,\theta)=-\frac{\mu s_{2 \theta}^2}{4 \a \b \left(-x+\a c_{ \theta}^2+\b s_{ \theta}^2\right)^2}~.
\een
We use the above expressions to write the equations of motion and the Virasoro constraints in a more compact form.

\section*{Acknowledgements}

I would like to thank Chong-Sun Chu for many discussions and for useful comments on the manuscript. I
would also like to thank Sam Bilson and George Georgiou  for useful comments on
the manuscript and George Zoupanos and the Department of Physics, of N. T. U. of Athens
for hospitality during the final stage of this work.
The research of the author is supported by an EPSRC studentship.

\end{document}